# Magnetic switching in granular FePt layers promoted by near-field laser enhancement

*Patrick W. Granitzka [‡,1,2], Emmanuelle Jal [‡,1,†,*], Loïc Le Guyader [1,12], Matteo Savoini [3], Daniel J. Higley [1,4,7], Tianmin Liu [1,5], Zhao Chen [1,5], Tyler Chase [1,4], Hendrik Ohldag [6], Georgi L. Dakovski [7], William Schlotter [7], Sebastian Carron [7,††], Matthias Hoffman [7], Padraic Shafer [8], Elke Arenholz [8], Olav Hellwig [9, †††], Virat Mehta [9, ††††], Yukiko K. Takahashi [10], J. Wang [10], Eric E. Fullerton [11], Joachim Stöhr [1], Alexander H. Reid [1] and Hermann A. Dürr [1*].*

[1] Stanford Institute for Materials and Energy Sciences, SLAC National Accelerator Laboratory, 2575 Sand Hill Road, Menlo Park, CA 94025, USA.

[2] Van der Waals-Zeeman Institute, University of Amsterdam, 1018XE Amsterdam, The Netherlands

[3] Institute for Quantum Electronics, Eidgenössische Technische Hochschule (ETH) Zürich, Auguste-Piccard-Hof 1, 8093 Zürich, Switzerland

[4] Department of Applied Physics, Stanford University, Stanford, California 94305, USA

[5] Department of Physics, Stanford University, Stanford, California 94305, USA

[6] Stanford Synchrotron Radiation Laboratory, SLAC National Accelerator Laboratory, 2575 Sand Hill Road, Menlo Park, CA 94025, USA.

[7] Linac Coherent Light Source, SLAC National Accelerator Laboratory, 2575 Sand Hill Road, Menlo Park, CA 94025, USA.

[8] Advanced Light Source, Lawrence Berkeley National Laboratory, Berkeley, California 94720, USA

[9] San Jose Research Center, HGST a Western Digital company, 3403 Yerba Buena Road, San Jose, California 95135, USA

[10] Magnetic Materials Unit, National Institute for Materials Science, Tsukuba 305-0047, Japan

[11] Center for Memory and Recording Research, UC San Diego, 9500 Gilman Drive, La Jolla, CA, 92093-0401, USA

[12] Spectroscopy & Coherent Scattering Instrument, European XFEL GmbH, Holzkoppel 4, 22869 Schenefeld, Germany

KEYWORDS: FePt nanoparticles, Ultrafast magnetism, X-ray scattering

ABSTRACT: Light-matter interaction at the nanoscale in magnetic materials is a topic of intense research in view of potential applications in next-generation high-density magnetic recording. Laser-assisted switching provides a pathway for overcoming the material constraints of high-anisotropy and high-packing density media, though much about the dynamics of the switching process remains unexplored. We use ultrafast small-angle x-ray scattering at an x-ray free-electron laser to probe the magnetic switching dynamics of FePt nanoparticles embedded in a carbon matrix following excitation by an optical femtosecond laser pulse. We observe that the combination of laser excitation and applied static magnetic field, one order of magnitude smaller than the coercive field, can overcome the magnetic anisotropy barrier between "up" and "down" magnetization, enabling magnetization switching. This magnetic switching is found to be inhomogeneous throughout the material, with some individual FePt nanoparticles neither switching nor demagnetizing. The origin of this behavior is identified as the near-field modification of the incident laser radiation around FePt nanoparticles. The fraction of not-switching nanoparticles is influenced by the heat flow between FePt and a heat-sink layer.

INTRODUCTION: Future magnetic data storage media will require magnetic nanoparticles with stable ferromagnetic order at diameters of only 10 nm and smaller[1]. In this respect, granular thin films of the $L1_0$-ordered phase of FePt displaying perpendicular magnetic anisotropy are one of the most suitable storage media. The FePt nanoparticles composing such granular materials remain ferromagnetic as a result of the strong magnetocrystalline anisotropy needed to overcome the superparamagnetic limit[2–5]. However, a byproduct of strong magnetocrystalline anisotropy is the large magnetic field required to reverse the nanoparticle magnetization. Applications strive to reduce the magnetic switching field by locally heating the nanoparticles above their Curie temperature with a laser in order to thermally assist the switching, a technique known as heat-assisted magnetic recording[6].

To date, the influence of the collective dielectric response of FePt nanoparticles on the magnetization switching has not been studied in detail. It is well known that the optical laser field can be dramatically enhanced via plasmonic resonances in the vicinity of metallic nanostructures such as Au[7] and Ag[8] nanosystems. Laser pulse shaping has been used to control dielectric and plasmonic responses in order to achieve sub-wavelength control of optical laser near fields[9,10]. Similar, albeit smaller, laser-field enhancements have been reported to occur near FePt nanoparticles[6,11]. This leads us to the obvious and important question: Do the dielectric properties of granular FePt layers affect the laser-assisted magnetic switching of these materials? To address this question, we study the well-established ultrafast demagnetization of FePt nanoparticles after a femtosecond (fs) optical excitation[12,13] to disentangle the spatially varying response of individual nanoparticles. Contrary to the heat-assisted magnetic recording process, any heating effects introduced by the fs excitation here does not heat up the FePt nanoparticles above their Curie temperature[12]. Time-domain measurements can then distinguish magnetic switching during or immediately after laser excitation as observed for all-optical switching[14] and precessional switching in applied magnetic fields on much slower time scales[13]. We employ time-resolved magnetic small-angle x-ray scattering to show that there is a reproducible switching of a large fraction of the illuminated FePt nanoparticles due to the near-field modifications of the incident laser pulses by neighboring nanoparticles. We quantify the amount of not-switching FePt nanoparticles and demonstrate that the switching probability is enhanced by an increased latency of the deposited laser-energy before being transported to a heat sink. We note that our results are of importance for a microscopic understanding of the recently observed so-called all-optical magnetic switching in FePt granular films[14–21].

X-RAY RESONANT MAGNETIC SCATTERING: To follow the magnetism dynamics of granular thin FePt films in an out-of-plane applied magnetic field, we performed time-resolved small-angle x-ray scattering experiments with the Soft X-ray Materials Science (SXR) instrument of the Linac Coherent Light Source (LCLS) x-ray free-electron laser at the SLAC National Accelerator Laboratory. We used 1.5 eV ultrashort laser pulses as a pump, and ultrashort soft x-rays pulses in resonance with the $2p - 3d$ core-valence $L_3$ absorption edge of Fe as a probe (see methods). Laser-pump and x-ray-probe pulses arrive collinearly at normal incidence to the sample with a variable time delay (see Fig. 1a). The scattering pattern is recorded for each time delay for right circular polarized as well as left circular polarized x-rays and for magnetic fields of $\mu_0 H\pm = \pm 0.4$T applied along the x-ray incidence direction. When x-rays pass through a thin film, their transmittance becomes modulated by the spatially varying chemical and magnetic distribution in the sample. This leads to a characteristic far-field diffraction pattern on a detector behind the sample[22]. The diffraction pattern is given by the Fourier transform of the spatial variation of the atomic scattering factor. Usually this atomic scattering factor is proportional to the electronic charge and known as the Thomson scattering term. However, for a $3d$ transition metal such as Fe with the x-ray energy tuned to the $2p - 3d$ core-valence resonance ($L_3$ edge) the absorption also depends strongly on the magnetic state, i.e. the size of the magnetic moment and its orientation relative to the x-ray helicity[23]. The atomic scattering factors at resonance can be summarized by writing the scattering intensity for opposite x-ray helicities or opposite magnetization directions as[24,25]:

$$I_{q,\pm} = |C_q|^2 + |S_q|^2 \pm 2 \cdot \text{Re}[C_q^* S_q] \qquad \text{eq. (1)}$$

$C_q$ and $S_q$ represent the Fourier transforms at wavevector, q, of the charge and spin distribution throughout the sample. The $C_q S_q$ cross term is the most interesting since it allows us to assess the average magnetic moment, $S_q$, of all nanoparticles separated by a distance $2\pi/q$ from neighboring particles that contribute via charge scattering, $C_q$, to the cross term. A typical scattering pattern is shown in Fig. 1a. This rotationally symmetric diffraction pattern can be condensed to a 1D data set using an angular integration providing the intensity vs. wavevector transfer, q, in the sample plane (Fig. 1b and c).

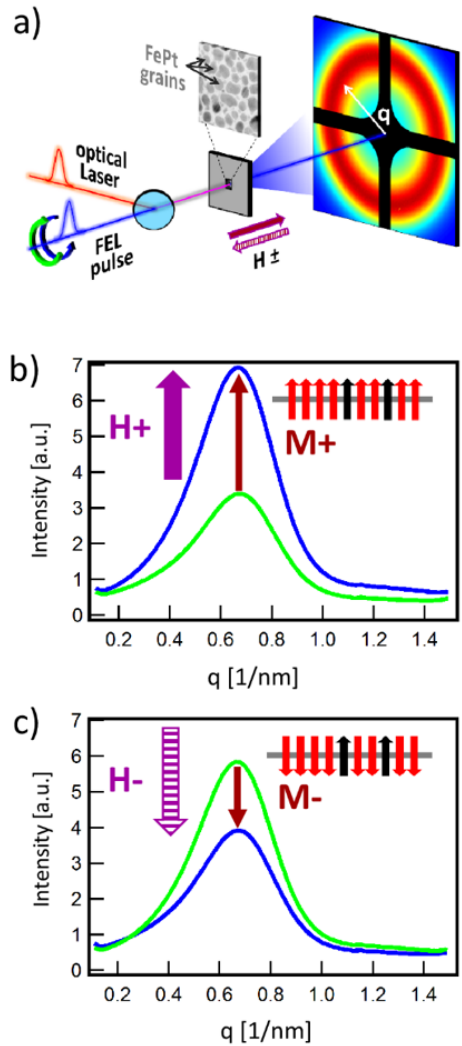

**Figure 1.** a) Schematic of the experimental set up described in the text. b) and c) show azimuthally integrated scattering intensities vs. wavevector, q, for supported FePt samples (see methods) ~$800\mu s$ after fs laser excitation measured with opposite x-ray helicities (blue and green curves) for positive and negative magnetic fields, respectively; insets: schematic of macro-spins of nanoparticles aligned to the respective applied H-fields (*red*: switching nanoparticles; *black*: not-switching nanoparticles).

RESULTS AND DISCUSSION:

*Identifying switching and not-switching nanoparticles:* Figures 1b and 1c show the scattering intensities before time zero as a function of the scattering vector, q, taken with opposite x-ray helicities for positive and negative applied magnetic fields, respectively. The sample, a FePt film supported by a membrane (see methods), was excited by fs laser pulses with a fluence of 11 mJ/cm$^2$. The x-ray pulses probe the sample ~800$\mu s$ after the fs optical laser excitation, well after the sample returned to equilibrium after Fe laser excitation and shortly before the subsequent pump-probe cycle. The peak visible in all four curves is due to diffraction from pairs of nanoparticles in the FePt films (see methods) and represents the average separation of $\frac{2\pi}{0.65 nm^{-1}} \approx$ 10 $nm$, in good agreement with the value obtained from electron microscopy images.

The dichroism, i.e. the difference in diffraction measured with opposite x-ray helicities, reflects the cross term, $C_q S_q$, of charge and spin scattering[19] from nanoparticle assemblies. If the magnetization, M, of all nanoparticles (marked by the dark red arrows in Fig. 1) followed the applied magnetic field, the scattering dichroism in Figs. 1b and 1c should be identical but of opposite sign. While the sign change is indeed observed, indicating that the majority of the nanoparticles reverse their magnetization, the observed dichroism magnitudes are different. With a positive H+ field (Fig. 1a) the absolute size of the dichroism is clearly larger than with negative H- field (Fig. 1b). The smaller size of the dichroism for negative H- field indicates that not all nanoparticles can reverse their magnetic moments and follow the applied magnetic field under the specific experimental conditions of fluence, pulse length and magnitude of the applied field. The situation is schematically depicted in the insets of Figs. 1b and 1c where each arrow represents the total magnetic moment of a nanoparticle. We can determine the fraction of switching nanoparticles that follow the applied field (red arrows) and the one of not-switching nanoparticles (black arrows) as being proportional to the difference and sum of the dichroism measurements, respectively, as shown in Figs. 1b and 1c. In our case, the magnetization that is switched represents ~80% of the total magnetization with the remaining not-switched magnetization contributing to ~20% (see Fig. 2a).

This behavior can be understood from the fact that in Fig. 1 the applied magnetic field is only $\pm 0.4\mu_0$T which is far below the coercive field of $2.8\mu_0$T required to magnetically switch half of the FePt nanoparticles at ambient temperature. We note that this is different to heat assisted magnetic recording where the sample is heated in near-equilibrium conditions above the Curie temperature[12,26] and then field-cooled in which case very small fields are sufficient to reverse the magnetization. Our case is characterized by non-equilibrium heating of electronic, spin and lattice degrees of freedom[12] as discussed in the following.

*Characterizing switching and not-switching nanoparticles:* Figures 2a and 2b show the azimuthally integrated scattering intensity vs. scattering vector q from switching (red curves) and not-switching (black curves) nanoparticles for FePt nanoparticle assemblies that are supported by a membrane or free-standing (see methods), respectively. In Figs. 2a and 2b the line shapes of the scattering intensity vs. wave vector, q, is largely identical for switching and not-switching nanoparticles, obtained from measurements such as shown in Fig. 1. This indicates that the spatial distribution of both switching and not-switching nanoparticles is very similar. Only at very small wave vectors in Fig. 2a do we observe a deviation. We note that in this q-range antiferromagnetic order between nanoparticles can cause additional contributions to the scattering signal that are not discussed here.

Figs. 2c and 2d display ultrafast demagnetization, a further characteristic of ferromagnetic materials exposed to intense fs-pulses of laser excitation[12,27]. Interestingly, only the switching nanoparticles demagnetize on sub-ps timescales. The extracted time constant of 146 ± 15 fs is in good agreement with literature values[12,13]. We observe demagnetization amplitudes of 72.2 ± 0.4% and 61.3 ± 0.8% in Figs. 2c and 2d, respectively. These values are in good agreement with the fluence dependent demagnetization behavior[12,27] due to the slightly different fs laser fluences of 11 mJ/cm² and 8 mJ/cm², respectively. In fact, we find no evidence that the different structure of supported and free-standing samples has any influence on the ultrafast demagnetization behavior. This demagnetization alone does not explain the magnetization reversal in an applied magnetic field significantly below the sample coercivity. Although it is not integral to the results of this paper, we mention that this study points to a laser-induced reduction of the magnetocrystalline anisotropy barrier between opposite FePt magnetization directions. However, a detailed discussion of this process is beyond the scope of the present paper.

The not-switching nanoparticles show a negligible amount of demagnetization indicating that these nanoparticles are significantly less exposed to fs optical laser radiation compared to their switching counterparts. We ascribe this behavior to a near-field laser modification around individual nanoparticles, which is discussed further in the next section. In contrast to the demagnetization behavior, there is a pronounced difference in the fraction of switching and not-switching nanoparticles for the two different sample types (supported and free-standing) shown in Fig. 2. While for the supported FePt sample in Fig. 2a not-switching nanoparticles constitute about 20% of all nanoparticles, the free-standing FePt sample used in Fig. 2b displays only 10% not-switching spins. It appears that these differences are directly related to the cooling of the nanoparticle assemblies on much longer times than the ones shown in Fig. 2.

Interestingly the supported film exposed to a higher fluence exhibits a larger percentage of not-switching spins (as shown in Fig. 2c) than the freestanding film (shown in Fig. 2d). One significant difference between those two samples is the existence of a heat sink layer underneath the FePt for the supported film (see methods). The presence of this heat sink leads to an efficient flow of deposited laser energy away from the FePt layer. In contrast, the heat dissipation in the freestanding FePt film occurs primarily laterally. The laser illuminated area is of the order of ~100 µm which implies that such lateral heat diffusion is much slower than vertical heat flow to a heat sink only several nm away. This provides a clear indication that it is the latency of deposited laser energy within the FePt nanoparticles that facilitates magnetic switching at sub-coercive magnetic field strengths in our experiments.

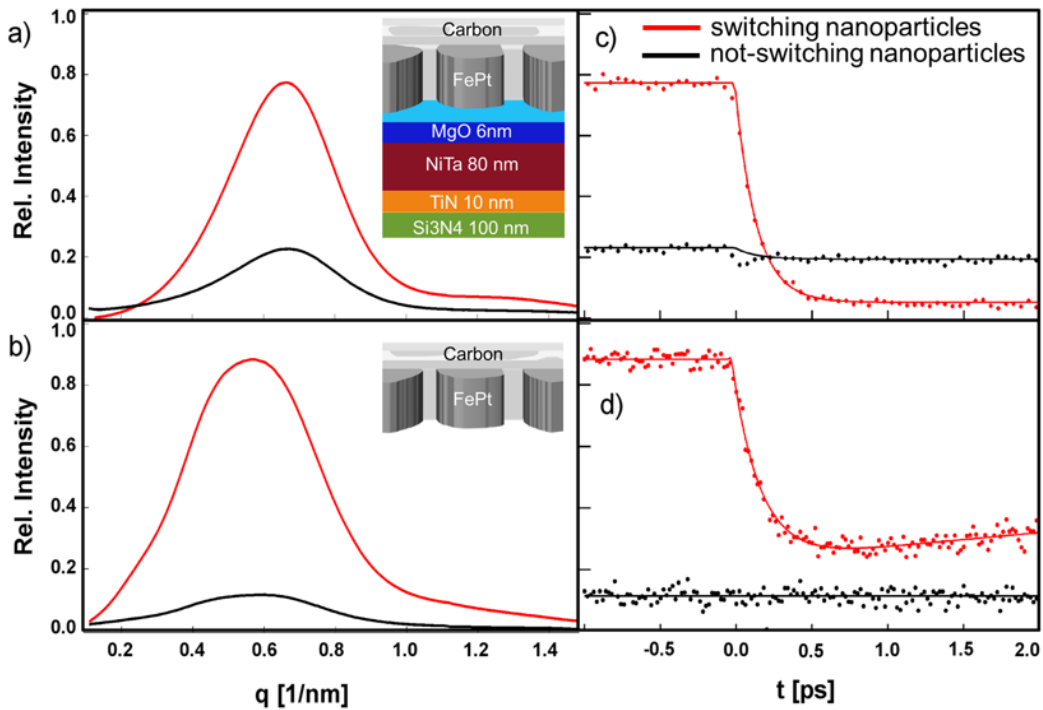

**Figure 2.** a) and b) display the radial diffraction intensity vs. wavevector, q. c) and d) display the time evolution of the peak maxima in a) and b), respectively. Red curves show the switching

nanoparticles whereas the black curves represent the not-switching spins obtained as described in the text. Data in a) and c) were obtained for supported FePt granular films pumped with 11 mJ/cm$^2$ while b) and d) are for free-standing FePt granular films pumped with 8 mJ/cm$^2$ (see methods). Insets show the composition of each samples.

*Modeling the near-field nanoparticle response of granular FePt films*: We modeled the spatial variation of the fs laser field in granular FePt films due to the dielectric properties of individual nanoparticles (see methods). This method has already proven to be an effective tool to calculate the spatial distribution of the deposited energy in relation to subsequent magnetization dynamics in various systems[7,28–31]. Individual nanoparticles are surrounded by a near-field modification of the incident laser radiation, as shown in Fig. 3a. The laser field is enhanced along the electric polarization direction (red) and reduced along the perpendicular direction (dark blue) compared to the incident laser radiation (light blue). If nanoparticles are close enough to one another, each experiences the modified laser field of its neighbor. This leads to changes in the optical absorption within the particles themselves as shown in Fig. 3b. It is clear from this figure that there is a pronounced spatial variation of the laser field leading to a spatially varying amount of absorbed laser energy. While increased absorption leads to ultrafast demagnetization in most nanoparticles, some also experience a reduced optical absorption. The latter would correspond to nanoparticles that are not-switching in Fig. 2. We can assign their spatial positions in Fig. 3b by introducing a cutoff optical absorption (see Fig. 3c). This was chosen in Fig 3c so that the ratio of nanoparticles above/below the cutoff matches the observed ratio of switching/not-switching spins in Fig. 2a. Fig. 3d summarizes the spatial distribution of not-switching (black) nanoparticles. It is evident that these correspond to more isolated nanoparticles. This isolation can manifest as a larger gap between the edges of neighboring FePt grains without necessarily increasing the center-to-center intergranular distance, which is consistent with the similar size distributions observed in either Fig. 2a or 2b. This is easy to understand as the near-field enhancement will decrease exponentially with distance[32] and therefore nanoparticles that are on average farther away from their neighbors experience less field enhancement than those that are statistically closer, and therefore accumulate less laser energy.

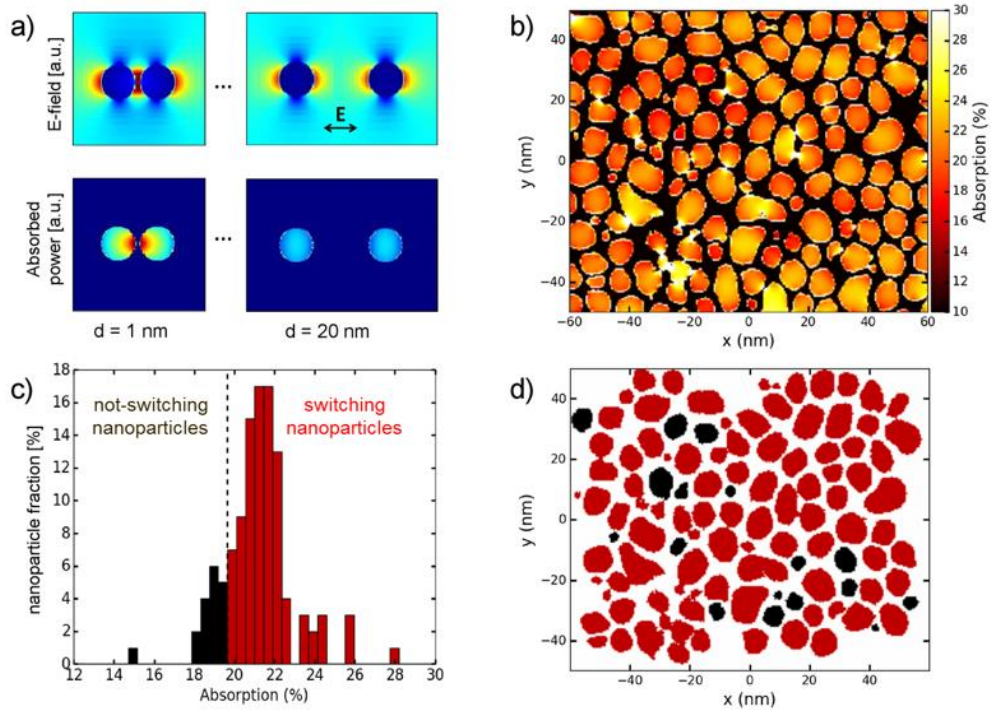

**Figure 3.** a) Calculated near-field modification of incident optical laser radiation surrounding pairs of circular FePt nanoparticles (6 nm diameter) for center-to-center distance of 1 and 20 nm. The color scale indicates reduction (dark blue) and increase (red) compared to the incident laser field (light blue). The linear electric field polarization of the incident radiation is indicated by the arrow. b) Calculated optical absorption (see methods) in a FePt granular film with the FePt nanoparticles separated by amorphous carbon (black areas). Circularly polarized excitation was used in the simulation to match the experiment. The nanoparticle distribution was taken from electron microscopy images. c) Distribution of the fraction of (not-)switching nanoparticles vs. absorbed fluence. The separation between switching and not-switching nanoparticles was chosen to match the measured values in Fig. 2a. d) FePt granular film as in b) but color coded according to switching and not-switching nanoparticles as shown in c).

CONCLUSION: We have shown that granular FePt films exhibit an interesting complexity when optically excited. Illumination with fs laser pulses leads to heat-assisted magnetic switching of the majority of the FePt nanoparticles. However, we observe that the magnetization of a significant fraction of nanoparticles (10-20%) cannot be reversed under these conditions as they experience a significantly reduced absorption of laser radiation. This is explained taking into account the spatially varying laser absorption in nanoparticles in the enhanced near-fields of adjacent nanoparticles. In addition, magnetization switching is found to depend sensitively on the retention time of deposited laser energy within the nanoparticles. The presence of a heat sink layer as is common in heat-assisted magnetic recording media improves the heat flow out of the nanoparticles, but results in larger ratios of not-switching to switching nanoparticles.

METHODS.

*SAMPLE FABRICATION.* The freestanding membrane was a single crystalline $L1_0$ FePt-C grown epitaxially onto a single-crystal MgO(001) substrate by compositionally graded sputtering deposition method[33] using Fe, Pt and C targets[18]. This process resulted in FePt nanoparticles of approximately cylindrical shape with heights of 10 nm and diameters in the range of 8-24 nm, with an average of 13 nm. The FePt nanoparticles form with a and b crystallographic directions along the MgO surface. The space in-between the nanoparticles is filled with amorphous carbon, which makes up 30% of the film's volume. Following the sputtering, the MgO substrate was chemically removed and the FePt-C films were floated onto copper wire mesh grids with 100 μm wide openings. The individual nanoparticles remained aligned during this process, as the particles are held in place by the carbon matrix[34].

The supported FePt granular layer was directly grown on a SiN membrane with a corresponding optimized underlayer seed layer structure to support the high temperature FePt growth on these substrates. We used a NiTa(80nm) heatsink layer, but had to separate it from the SiN(100nm) membrane with a TiN(10nm) barrier layer in order to avoid inter-diffusion between the NiTa and SiN as observed in previous generations of FePt membrane samples, which led to significant surface roughness enhancement. On top of the NiTa heatsink we deposit a MgO(6nm) seed layer that is out-of-plane textured in (001) direction with a mosaic crystallite spread of about 5-10 degrees in out-of-plane direction. The FePtC media layer was sputter deposited at about 650°C from a composite target with ~35vol% Carbon content with a nominal thickness of 7.8 nm in order to avoid second layer grain formation. All depositions were done at HGST, a Western Digital Company using a high-throughput multi-chamber industrial tool based on the Intevac Lean 200 platform. The layer structure was finally capped with a 3nm Carbon layer at RT to avoid any oxidation or corrosion. Average lateral grain size for this film was about 10 nm, so slightly smaller than for the other sample, and consistent with the scattering profiles in Fig. 2a and 2b

Before the time resolved x-ray experiment the samples were characterized by x-ray spectroscopy and scattering at beamline 4.0.2 of the Advanced Light Source in Berkeley. Then the samples were magnetically saturated in a $+7\mu_0$T field aligning all the spins and so nanoparticles into the up (+) direction.

*EXPERIMENTAL SET UP.* Time-resolved X-ray scattering measurements were performed in transmission mode in a collinear pump–probe geometry. Samples were photo-excited by circularly polarized 30 fs optical laser pulses with a central wavelength of 800 nm. The Fe charge and magnetic scattering was probed by 60fs circularly polarized x-ray pulses at the Fe $L_3$ absorption edge (706.8eV photon energy). Linearly polarized x-ray pulses from the LCLS x-ray free electron laser were passed through a Fe magnetic film to generate the circular polarization required for our experiment[25]. Scattered x-rays were recorded by a pn-CCD detector at repetition rate of 120 Hz. A constant magnetic field of $\pm 0.4T$ was applied during data acquisition. Measurements for opposite applied magnetic field orientations and x-ray helicities allowed us to separate the different scattering contributions in eq. (1) as described in the text. The x-ray wavelength at the $L_3$ edge of Fe is 1.7 nm and it is therefore possible to measure the inter-grain correlation length, meaning that the diffraction pattern mainly reflects the nanoparticle to nanoparticle distance in the sample plane as well as the magnetic correlation between FePt nanoparticles.

*OPTICAL SIMULATION.* The simulations are finite difference time domain (FDTD) simulations performed with the commercial software Lumerical FDTD[35]. The simulation consists of an area of 0.13 x 0.1 x 1 $\mu m^3$ with a nonuniform meshing with the smallest mesh cell being 0.5x0.5x0.5 $nm^3$ at the magnetic layer position. The boundary conditions are periodic in x and y to simulate an infinite sample. Along the z axis a perfectly matched layer is chosen to reduce unphysical reflections and minimize simulation time. The simulated FePt size, shape and distribution is obtained by importing scanning electron microscopy images acquired on the real sample. The shape and size is constant throughout the sample thickness. The FePt particles are embedded in a C-matrix. The dielectric constants used for the different materials are $\varepsilon = 3.9731 + i\ 17.358$ for FePt and $\varepsilon = 3.2396 + i\ 0.072$ for the C-matrix[36]. A set of two plane wave sources at a wavelength of 800 nm properly polarized and dephased is used to simulate the circularly polarized excitation. The light absorption $\frac{4\pi n k}{\lambda}|\vec{E}|^2$ [37], where $\vec{E}$ is the light electric field, is integrated throughout the FePt particles thickness. A good convergence of the simulations was obtained with variable time steps <0.1 fs and a total simulation time of ~50 fs, while the Fourier-transform-limited laser pulse was <10 fs long.

AUTHOR INFORMATION

Corresponding Author

*E-mail: hdurr@slac.stanford.edu.

*E-mail : emmanuelle.jal@gmail.com.

Present Addresses

† Sorbone Universitées, UPMC Univ Paris 06, UMR 7614, LCPMR, 75005 Paris, France ; and CNRS, UMR 7614, LCPMR,75005 Paris, France

†† California Lutheran University, 60 West Olsen Rd, Thousand Oak, CA 91360, U.S.A

††† Institute of Physics, Chemnitz University of Technology, Reichenhainer Straße 70, D-09107 Chemnitz, Germany and Institute of Ion Beam Physics and Materials Research, Helmholtz-Zentrum Dresden–Rossendorf, 01328 Dresden, Germany

†††† Thomas J. Watson Research Center, 1101 Kitchawan Road, Yorktown Heights, New York 10598, USA

Author Contributions

The manuscript was written through contributions of all authors. All authors have given approval to the final version of the manuscript. ‡These authors contributed equally. E.J., P.G., and H.A.D coordinated work on the paper with contributions from M.S., L.LG., O.H., and A.H.R and discussion with all authors. P.G., E.J., D.J.H., T.L., Z.C., T.C., H.O., Y.K.T., H.A.D., and A.H.R performed the x-ray diffraction measurements. G.L.D. and W.S. operated the SXR beamline, S.C. provided and operated the pn-CCD, and M.H. operated the pump laser and synchronization. P.S. and E.A. operated the beamline 4.0.2 of the Advanced Light Source in Berkeley. E.J., P.G., L.LG., S.C., A.H.R and H.A.D performed the data analysis. O.H., V.M., Y.K.T., and E.F grew the samples, M.S. and S.L.J. made the optical simulation. A.H.R., E.F., J.S. and H.A.D, designed and coordinated the project.

Funding Sources

Work at SIMES is supported by the Department of Energy, Office of Science, Basic Energy Sciences, Materials Sciences and Engineering Division, under Contract No. DE-AC02-76SF00515. Work at UCSD is supported by the Office of Naval Research MURI program. Use of the Linac Coherent Light Source, SLAC National Accelerator Laboratory, is supported by the U.S. Department of Energy, Office of Science, Office of Basic Energy Sciences under Contract No. DE-AC02-76SF00515.

ACKNOWLEDGMENT

We are grateful to Chris O'Grady for his help and discussion about the data analysis. L.L.G. thanks the Volkswagen-Stiftung for financial support through the Peter-Paul-Ewald Fellowship.

For Table of Contents Only

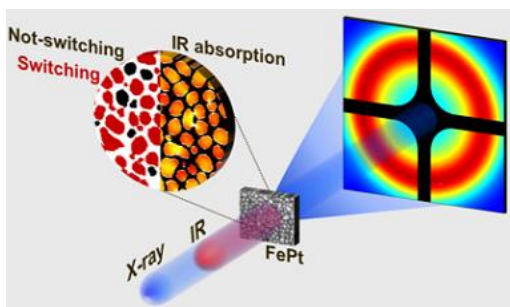